\documentclass[aps,prl,twocolumn,showpacs,floatfix]{revtex4}
\usepackage{graphics,amsmath,amssymb,amstext}
\usepackage{epsfig}
\usepackage{color}
\begin{document}

\title{Study of resonance states of $^{11}$Be with isospectral bound state microscopic potential}
\author{S. K. Dutta$^1$}
\author{D. Gupta$^2\footnote{Corresponding author email:
dhruba@jcbose.ac.in}$}
\author{D. Das$^2$}
\author{Swapan K. Saha$^2$}
\affiliation{$^1$Department of Physics, B.G. College, Berhampore,
Murshidabad 742101, India} \affiliation{$^2$ Department of Physics,
Bose Institute, 93/1 A.P.C. Road, Kolkata 700009, India}
\date{\today}

\begin{abstract}
The theoretical procedure of supersymmetric quantum mechanics (SQM)
is adopted for the first time to study quasi-bound states of a
weakly bound nuclear system using microscopic potential. The density
dependent M3Y (DDM3Y) effective interaction was found earlier to
give a satisfactory description of radioactivity, nuclear matter and
scattering. In the present work, we have generated a two-body
potential microscopically in a single folding model using the DDM3Y
effective interaction. From this potential, SQM generated a family
of isospectral potentials for $^{11}$Be ($^{10}$Be + n). We
investigated the 5/2$^+$, 3/2$^-$ and 3/2$^+$ resonance states of
$^{11}$Be. The experimental data and the present calculations of
excitation energies of the above resonance states are found to be in good agreement.\\


\end{abstract}

\maketitle

\section {Introduction}

\noindent The dynamics of  drip line nuclei are quite different from
those of stable nuclei~\cite{TA88}. These nuclei present a challenge
for the existing nuclear theory~\cite{CS94,AG71,SU91,SU88} to
provide an appropriate framework, within which it would be possible
to account for the peculiar properties of nuclei far off the
stability line. New theoretical tools are needed for a complete and
consistent description of such nuclei.\\

\noindent We have undertaken the task of studying the weakly bound
nucleus $^{11}$Be which has a spectrum of resonant
states~\cite{SH04}. Conclusive information about its resonant states
became available with the advent of highly sophisticated radioactive
ion beam facilities. These metastable states are sources of
information regarding nucleon-nucleon interaction at unusually large
distances in a low nuclear density medium.  The $^{11}$Be nucleus
has two bound states, the ${1/2}^+$ ground state and the ${1/2}^-$
excited state. All other excited states are the resonant states.
Therefore, investigation of resonant states becomes a necessity due
to insufficient number of bound states. Analysis of spectra of
resonant states of $^{11}$Be provides valuable information about the
nature of interactions among the constituents of the system. For
$^{11}$Be nucleus having a well known halo structure~\cite{CA01,
CR11, MO12}, it is extremely important if we can extract any
information regarding its wavefunction from resonant state analysis.
Thus, using the present theoretical tool, unknown
resonance states of newly found exotic nuclei can be studied.\\

\noindent In the present work we found that a two-body model gives a
good description of $^{11}$Be, in which the valence nucleon moves
independently in a mean field. We have studied weakly bound
$^{11}$Be nucleus in the framework of a two-body model
consisting of an inert core of $^{10}$Be and a single valence neutron.\\

\section {Theory}

\noindent We adopt an efficient theoretical
technique~\cite{SD03,SD04,SD042} for the resonant states, which
treats the ground and resonant states on the same footing. Our
calculational procedure circumvents the numerical difficulties posed
by such low lying broad resonances. We use the same interaction for
the study of both ground as well as resonant states. The two body
equation is solved to obtain an effective potential, $v(r)$, as a
function of the radial length variable $r$, corresponding to the
$J^{\pi}$ quantum number of the ground state. The effective
potential $v(r)$ also includes the centrifugal barrier. Solution of
the radial equation with this potential gives the ground state wave
function $\psi_{0}$. We then use the prescription of supersymmetric
quantum mechanics (SSQM)~\cite{ssqm,khr,nieto} to calculate a
potential ($\hat{v}$) which is strictly isospectral with $v$, but
has properties desirable for an accurate calculation of the resonant
state having the same $J^{\pi}$. However this procedure does not
apply if the ground state is unbound. Thus it cannot be applied to a
resonant state, corresponding to which no bound state having the
same $J^{\pi}$ exists. For such resonances, we adopt a slightly
different method, using a lesser known result of SSQM, {\it viz.}, a
bound state in the continuum (BIC); one can construct an isospectral
potential (IP) which has a normalizable positive energy solution at
a selected energy. Once again this IP has desirable properties which
can be utilized for an accurate calculation of such resonances. The
effective two body potential thus obtained is judiciously used to
extract important information about resonant states. The potential
$v(r)$, in general, presents a shallow well followed by a low and
wide barrier as  $r$ increases. For a finite barrier height, a
system may temporarily be trapped inside the shallow well, when its
energy is close to the resonance energy. So, in principle, one can
find quasi-bound states in such a shallow potential. However there
is a large probability for tunneling through the barrier which gives
rise to broad resonance widths. Calculation of resonance energies of
weakly bound unstable nuclei thus presents a very challenging
problem. Our technique as described above circumvents this
difficulty and obtain resonance energies in a precise manner.\\

\begin{figure}
\centering
\includegraphics[scale=0.5]{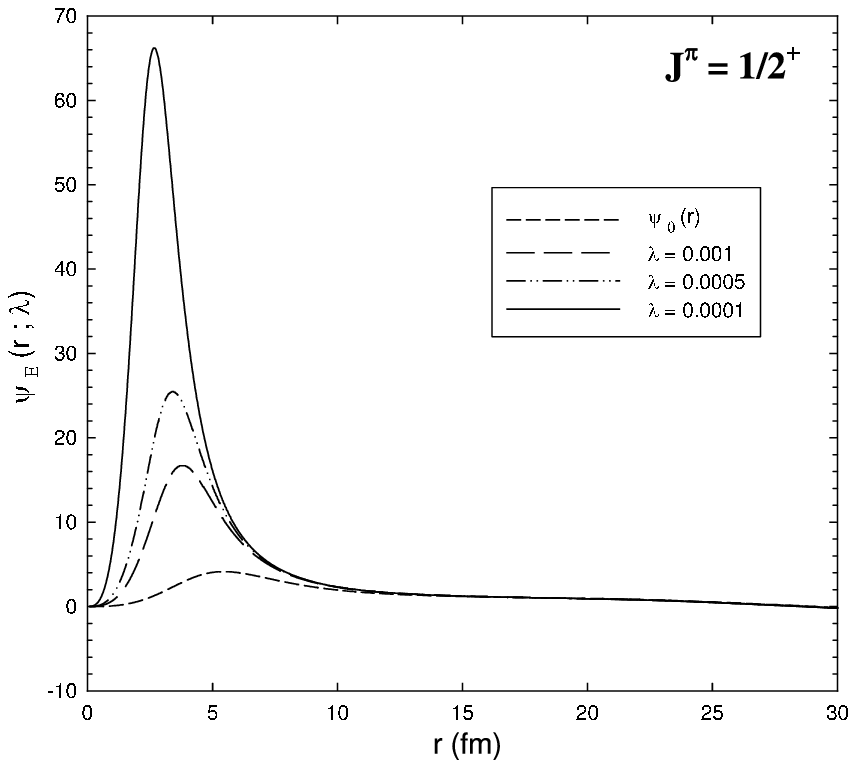}
\caption{Wave functions for the ground state ($\psi_{0}(r)$) of the
original potential $v(r)$ and isospectral potentials for $\lambda =
10^{-3}, 5 \times 10^{-4}$ and $10^{-4}$ for the ${1/2}^+$ state of
$^{11}Be$.} \label{fig1}
\end{figure}

\noindent From Darboux transformation of SSQM~\cite{ssqm,dbx,keu},
one can construct a partner potential for any given potential, such
that the spectrum of the partner is identical to that of the
original, except that the ground state of the original is missing in
the spectrum of the partner potential. Using this procedure for
deleting and then subsequently reinstating the ground state of a
potential $v(r)$, one can construct a family of strictly isospectral
potentials (IP) $\hat{v}(r,\lambda)$, involving an arbitrary
parameter $\lambda$. It is observed that as $\lambda$ $\rightarrow$
0+, $\hat{v}(r,\lambda)$ develops a narrow and deep attractive well
followed by a high barrier near the origin. Since $\hat{v}$ is
strictly isospectral with $v$, the energy $E_{R}$ at which resonance
occurs in $\hat{v}$ is exactly the same as that in $v$, but the
system has a much higher probability of being trapped in a deep well
followed by the high barrier of $\hat{v}$. Thus, while the resonance
at an energy $E_{R}$ is very broad in $v(r)$, it is very sharp in
$\hat{v}$, facilitating calculation of $E_{R}$ and the width
$\Gamma$. However this procedure needs a bound ground state in
$v(r)$. For many of the resonances in weakly bound halo nuclei,
there are no bound states with the same $J^{\pi}$ quantum numbers.
For such resonances the above procedure of using isospectral
potentials does not apply. The idea of isospectral potential has
been extended by Pappademos et al.~\cite{pappa} to scattering states
with positive energy in the continuum. While the wave functions in
the continuum are non-normalizable, following Pappademos et al. one
can construct a normalizable wave function at a selected energy -
thus representing a bound state in the continuum (BIC). The BIC is a solution of the 
equation with an isospectral potential $\hat{v}(r,\lambda)$. As the
theory predicts, it is found in practice that resonance energy does
not depend on the choice of $\lambda$. So $\lambda$ is suitably
chosen to optimize the stability of the resonant state. It again
preserves the spectrum of the original potential, only it adds a
discrete BIC at a selected energy. In the present work we have shown
that application of SSQM formalism together with BIC technique
describes resonance states of $^{11}$Be very well. Being a bound
state in the continuum of the isospectral family, this technique
requires less numerical effort and offers better precision.\\

\begin{figure}
\centering
\includegraphics[scale=0.5]{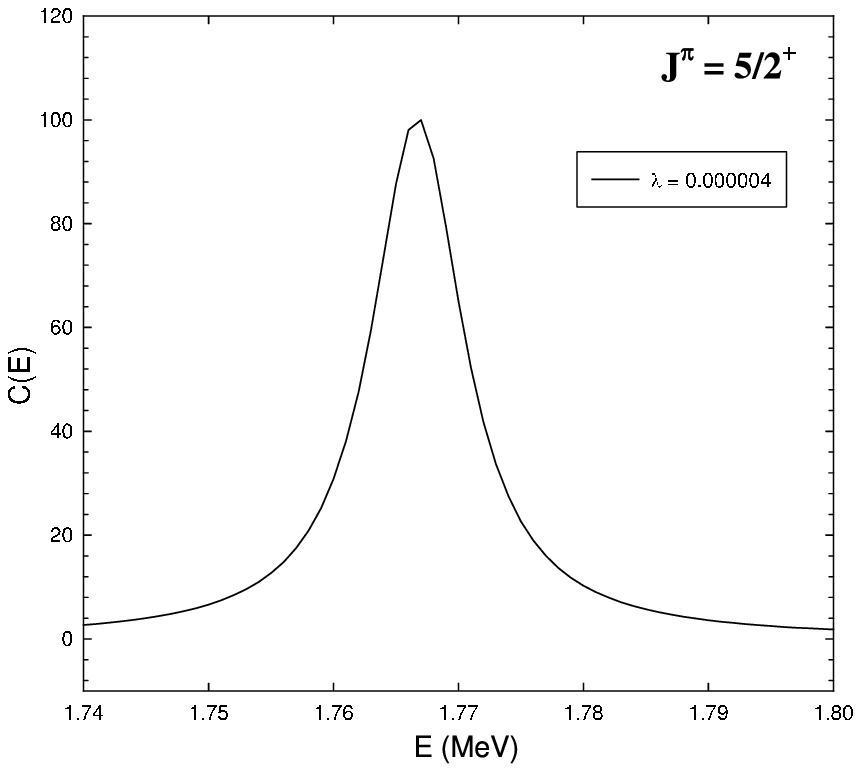}
\caption{Probability $C(E)$ as a function of energy $E$ for $\lambda
= 4 \times 10^{-5}$ for the ${5/2}^+$ state of $^{11}Be$.}
\label{fig2}
\end{figure}

\begin{figure}
\centering
\includegraphics[scale=0.5]{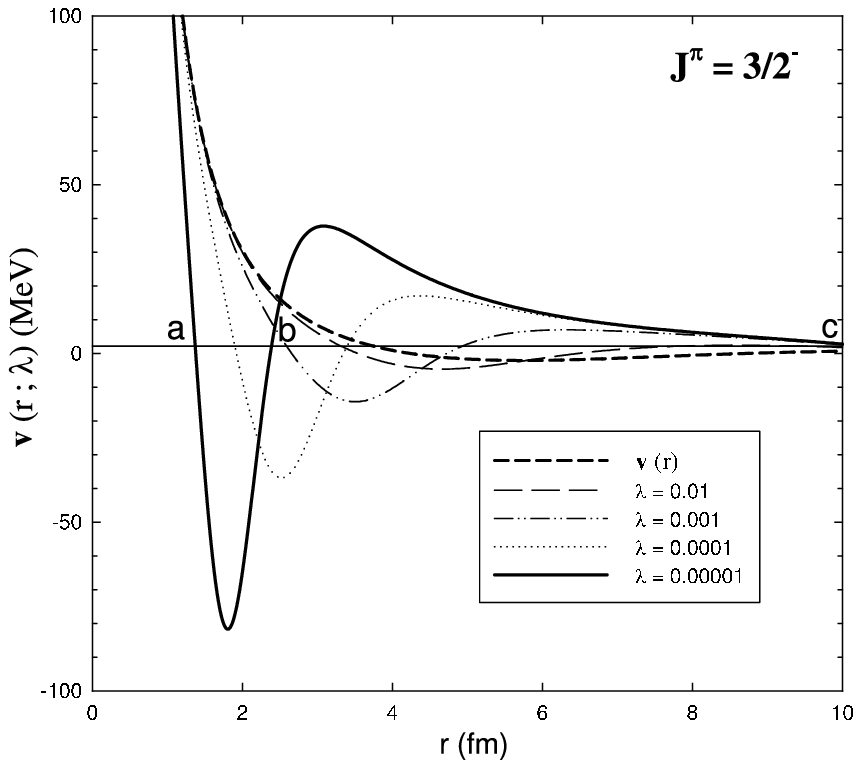}
\caption{The original folding potential $v(r)$ along with one
parameter family of isospectral potentials $\hat v(r;\lambda)$ for
$\lambda = 10^{-2}, 10^{-3}, 10^{-4}$ and $10^{-5}$ for the
${3/2}^-$ state of $^{11}Be$. The turning points (a,b,c) are
marked.} \label{fig3}
\end{figure}

\subsection {Single folding potential {\it v(r)}}

The nucleon-nucleus potential is obtained in a single folding
calculation by using the density distribution of the nucleus and the
nucleon-nucleon effective interaction~\cite{SA79} as,
\begin{equation}
U(\vec{r_1}) = \int \rho_2(\vec{r_2}) v(|\vec{r_1}-\vec{r_2}|)d^3\vec{r_2}\\
\end{equation}
where $\rho_2(\vec{r_2})$ is density of the nucleus at $\vec{r_2}$
and $v(r = |\vec{r_1}-\vec{r_2}|)$ is the effective interaction
between two nucleons at the sites $\vec{r_1}$ and $\vec{r_2}$. The
finite range M3Y effective interaction $v(r)$~\cite{r9}, is based
upon a realistic G-matrix and was constructed in an oscillator
basis. Since it is an average over a range of nuclear densities and
energies, it has no explicit dependence on density or energy. A weak
energy dependent effect is contained only in an approximate
treatment of single-nucleon knock-on exchange. The density and
energy averages are adequate at lower energies for the real part of
the heavy ion optical potentials. For scattering at higher energies,
explicit density dependence was introduced~\cite{r10,r11}. The
present calculations use this density dependent M3Y (DDM3Y)
effective NN interaction with an added zero-range pseudo potential
given by,

\begin{equation}
  v(r,\rho,E) = t^{\rm M3Y}(r,E)g(\rho,E)~~{\rm MeV}
\end{equation}
\noindent where $E$ is incident energy and

\begin{equation}
  t^{\rm M3Y} = \left[7999 \frac{e^{ - 4r}}{4r} - 2134\frac{e^{- 2.5r}}{2.5r}\right] {\rm MeV} + J_{00}(E)
  \delta(r)
\end{equation}
\noindent The zero-range pseudo-potential~\cite{r10} represents the
single-nucleon exchange term and is given by

\begin{equation}
 J_{00}(E) = -276 (1 - (0.005~{\rm MeV^{-1}})E/A)~~{\rm MeV.fm^3}
\end{equation}
\noindent while the density dependent part is taken~\cite{r11} to be

\begin{equation}
g(\rho, E) = c (1 - b(E)\rho^{2/3})
\end{equation}
\noindent taking care of the higher order exchange and Pauli
blocking effects. Here $\rho~=\rho_2$ is the spherical ground state
density of the nucleus. We use the ``realistic" neutron and proton
densities as described in eqn. (2) of ref.~\cite{KO97} for $^{10}$Be
nucleus. The constants of this interaction $c$ and $b$ when used in
single folding model description, are determined by nuclear matter
calculations~\cite{BA04} as 2.07 and 1.62 fm$^2$ respectively.
Earlier, the DDM3Y interaction is found to provide a unified
description of radioactivity, nuclear matter and
scattering~\cite{GU06,BASUproton}. Thus it is very appropriate to
study quasi-bound states of a weakly bound nuclear system using this interaction.\\

\begin{figure}
\centering
\includegraphics[scale=0.5]{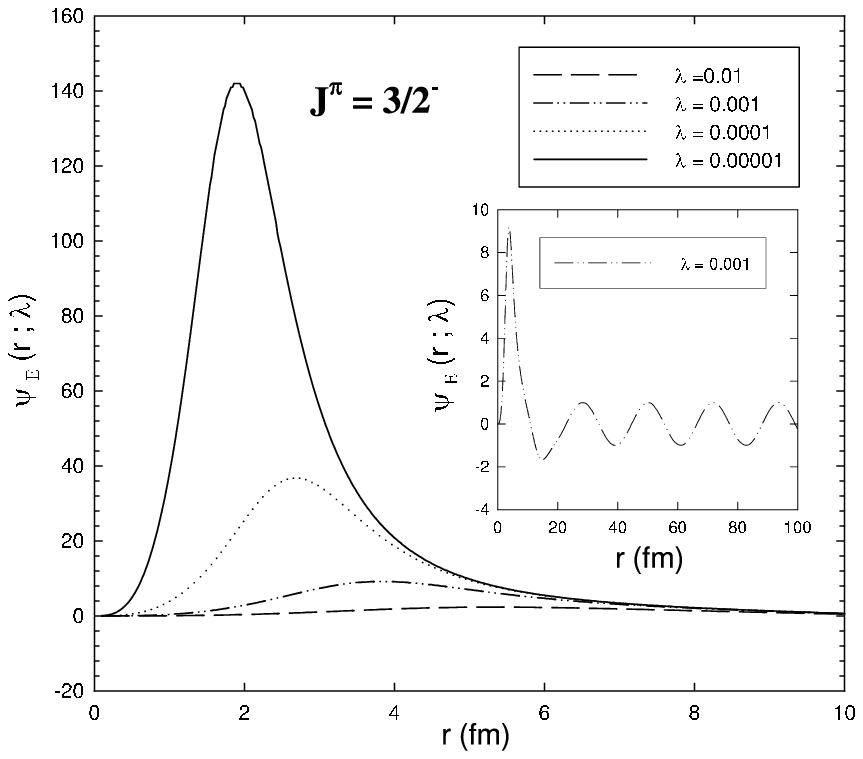}
\caption{Wave function (in arbitrary units) at the excitation energy
2.67 MeV for $\lambda = 10^{-2}, 10^{-3}, 10^{-4}$ and $10^{-5}$ for
the ${3/2}^-$ state of $^{11}Be$.} \label{fig4}
\end{figure}

\subsection {Isospectral potential of {\it v(r)}}

\noindent As mentioned earlier, for a particular $v(r)$, it is
possible to construct a family of strictly isospectral potentials
$\hat{v}(r;\lambda)$ depending on an arbitrarily chosen parameter
$\lambda$. This is done by the technique of SSQM, which is briefly
reviewed in the following paragraph. The equation for the ground
state wave function $\psi_{0}(r)$ is
\begin{equation}
\left(-\frac{\hbar^{2}}{2\mu}\frac{d^{2}}{dr^{2}} + v(r) - E_{0}
\right) \psi_{0}(r) = 0
\label{isoeq}
\end{equation}
 where $E_{0}$ is the energy of the ground state.
Shifting the energy scale, such that the ground state is at zero energy
(corresponding potential is denoted by $v_{1}(r)=v(r) - E_{0}$) one can
factorize the Hamiltonian as
\begin{equation}
H_{1}= -\frac{\hbar^2}{2\mu}\frac{d^{2}}{dr^{2}} + v_{1}(r) = A^{\dag}A
\end{equation}
where $ A = \displaystyle \frac{\hbar}{\sqrt{2\mu}}\frac{d}{dr} +
W(r)$, $A^{\dag} = -\displaystyle
\frac{\hbar}{\sqrt{2\mu}}\frac{d}{dr} + W(r)$ and the
`superpotential'~\cite{ssqm,nieto} $W(r)$ is given by
\begin{equation}
W(r) = -\frac{\hbar}{\sqrt{2\mu}}\frac{\psi_{0}^{\prime}(r)}{\psi_{0}(r)} \:.
\end{equation}
It can be easily verified that
\begin{equation}
v_{1}(r) = W^{2}(r) - \frac{\hbar}{\sqrt{2\mu}}W^{\prime}(r) \:,
\end{equation}
where a prime denotes differentiation with respect to the argument. Then
SSQM prescribes a partner Hamiltonian
$ H_{2} = AA^{\dag} = - \frac{\hbar^{2}}{2\mu}\frac{d^{2}}{dr^{2}} + v_{2}(r) $ corresponding
to a partner potential
\begin{equation}
 v_{2}(r) = W^{2}(r) + \frac{\hbar}{\sqrt{2\mu}}W^{\prime}(r).
\end{equation}
It can be shown that the Hamiltonians $H_{1}$ and $H_{2}$ have
identical spectra (but different eigenfunctions), except that the
ground state of $H_{1}$ is missing in the spectrum of $H_{2}$. To
reinstate this missing ground state (thus making the two potentials
strictly isospectral), Nieto obtained~\cite{nieto} the most general
superpotential, $\hat{W}$ such that
\begin{equation}
v_{2}(r) = \hat{W}^{2} + \frac{\hbar}{\sqrt{2\mu}}\hat{W}^{\prime}.
\end{equation}
For a given $v_{2}(r)$, above equation is a non-linear equation for
$\hat{W}(r)$; hence the solution is not unique and one can get a
one-parameter family of solutions for $\hat{W}$.
\begin{equation}
\hat{W}(r;\lambda)=W(r)+\frac{\hbar}{\sqrt{2\mu}}\frac{d}{dr}\ln[I_{0}(r) +
\lambda]
\end{equation}
with
\begin{equation}
I_{0}(r)={\displaystyle\int}_{0}^{r}
{[\psi_{0}(r^{\prime})]}^{2} dr^{\prime}.
\end{equation}
Here $\lambda$ is a real parameter, whose value can be in the range
\mbox{$0<\lambda<\infty$}. Since $\psi_{0}$ is normalized, $0\leq
I_{0}(r)\leq 1$ and therefore $\lambda \leq 0$ is not allowed. Since
$\hat{W}$ satisfies the most general supersymmetric (SUSY)
partner~\cite{nieto} of $v_{2}(r)$, it follows that
\begin{equation}
\begin{array}{ccl}
\hat{{v}}_{1}(r;\lambda)&=& \hat{W}^{2}(r;\lambda)-
\displaystyle \frac{\hbar}{\sqrt{2\mu}}\hat{W}^{\prime}(r;\lambda)\\
                           &=& v_{1}(r) - 2\displaystyle \frac{\hbar^{2}}{2\mu}
\frac{d^{2}}{dr^{2}}\ln[I_{0}(r) + \lambda] \:.
\end{array}
\label{isovhat}
\end{equation}
It is then strictly isospectral with $v_{1}(r)$. Although both have
identical spectra, the nature of $\hat{v}_{1}$ may be very different
from that of $v_{1}(r)$, depending on the value of $\lambda$. The
limits $\lambda \rightarrow\pm\infty$ correspond to the original
potential $v_{1}(r)$. For $\lambda\rightarrow0+$,
$\hat{v}(r;\lambda)$ develops a deep and narrow attractive well
followed by a high barrier, near the origin. Then the wave function
for the potential $\hat{v}(r;\lambda)$ develops a sharp peak near
the origin, for $\lambda\rightarrow 0+$. We utilize this feature to
calculate the resonance states  in the potential $v_{1}(r)$, or
equivalently $v(r)$. Wave functions corresponding to $v(r)$ and for
isospectral potentials for three different $\lambda$ values for the
${1/2}^+$ state of $^{11}Be$ are shown in Fig. 1.\\

\begin{figure}
\centering
\includegraphics[scale=0.5]{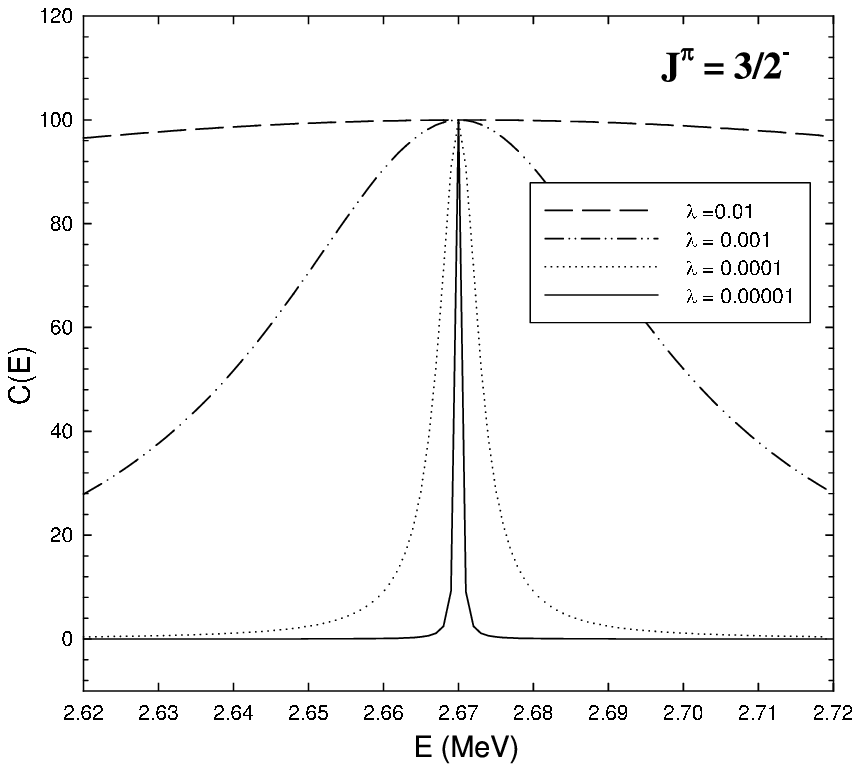}
\caption{Probability $C(E)$ as a function of energy $E$ for $\lambda
= 10^{-2}, 10^{-3}, 10^{-4}$ and $10^{-5}$ for the ${3/2}^-$ state
of $^{11}Be$.} \label{fig5}
\end{figure}

\subsection {Isospectral potential to locate a bound state in the continuum (BIC)}

\noindent In order to calculate the isospectral potential
$\hat{v}_{1}( r;\lambda)$ for a given $J^{\pi}$ one needs the
knowledge of the normalized ground state wave function $\psi_{0}(r)$
for the same $J^{\pi}$. However in  halo nuclei, at most only the
ground state is bound  and for resonances with $J^{\pi}$ different
from that of the ground state, we have no normalizable lowest energy
state of the same $J^{\pi}$. Thus the procedure above is not
applicable for such resonances. For such cases we adopt a somewhat
different approach. We follow Pappademos et al~\cite{pappa} and
generalize the above procedure starting with any solution
$\psi_{E}(r)$ of Eq.~(\ref{isoeq}) corresponding to energy $E>0$
(subject to the boundary condition that $\psi_{E}(0)=0$). Thus
$\psi_{E}$ is not square integrable, but it oscillates as $r$
increases. One can verify by direct substitution that

\begin{equation}
\hat{\psi}_{E}(r;\lambda) = \frac{\psi_{E}(r)}{I_{E}(r) + \lambda} \:,
\label{psihat}
\end{equation}
where
\begin{equation}
I_{E}(r) = {\displaystyle\int}_{0}^{r}{[\psi_{E}(r^{\prime} )]}^
{2}dr^{\prime}
\label{ie}
\end{equation}
satisfies Eq.~(\ref{isoeq}) with $v(r)$ replaced by

\begin{equation}
\hat{v}(r;\lambda) = v(r) - \frac{\hbar^2}{2\mu}\left[\frac{4\psi_{E}(r)\psi_{E}^
{\prime}(r)}{I_{E}(r) + \lambda}
+\frac{2(\psi_{E}(r))^{4}}{{(I_{E}(r) + \lambda)}^{2}}\right] \:,
\label{vhat}
\end{equation}
when $\psi_{E}(r)$ satisfies Eq.~(\ref{isoeq}). Thus for the
potential $\hat{v}(r;\lambda)$ given by Eq.~(\ref{vhat}), which
depends on the arbitrary parameter $\lambda$, $\hat{\psi}_{E}(r)$ is
the solution corresponding to the same energy $E$. However since
$\psi_{E}(r)$ oscillates with a constant amplitude in the asymptotic
region, $I_{E}(r)$ increases approximately linearly with $r$ for
large $r$ and from Eq.~(\ref{psihat}) we see that
$\hat{\psi}_{E}(r;\lambda)$ is normalizable. Thus
$\hat{\psi}_{E}(r;\lambda)$ represents a bound state in the
continuum (BIC) of $\hat{v}(r;\lambda)$, which is isospectral with
$v(r)$. Since $0\leq I_{E}(r)<\infty$, negative values of $\lambda$
are no longer allowed. As before, $\hat{v}(r;\lambda)$ develops a
deep and narrow well followed by a high barrier near the origin for
$\lambda\rightarrow0+$ and
 approaches  $v(r)$ for $\lambda\rightarrow+\infty$.\\

\begin{figure}
\centering
\includegraphics[scale=0.5]{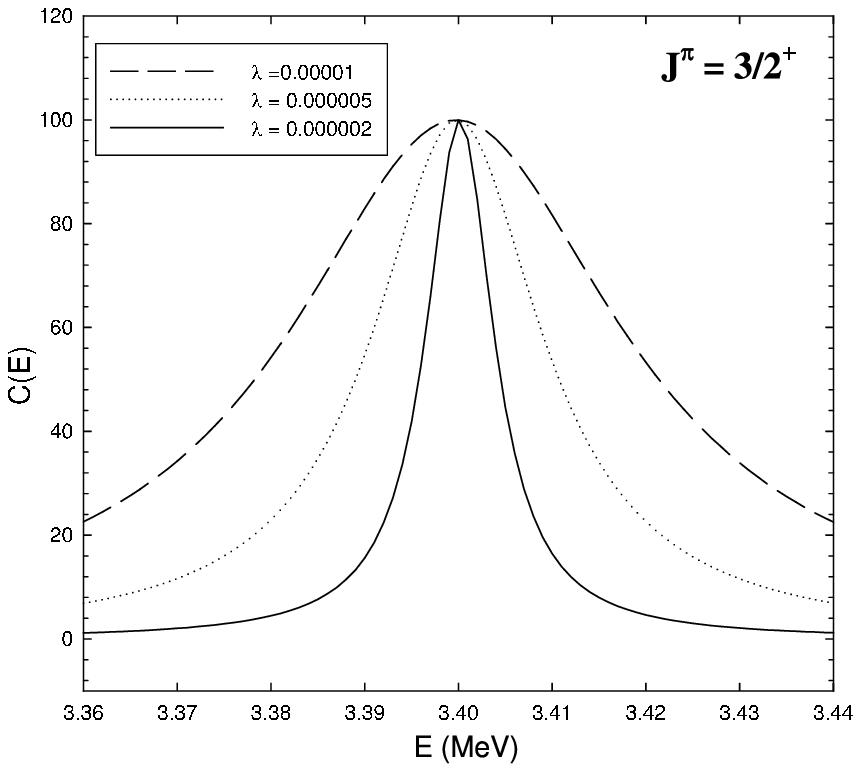}
\caption{Probability $C(E)$ as a function of energy $E$ for $\lambda
= 1.0 \times 10^{-5}, 5.0 \times 10^{-6}$ and $2.0 \times 10^{-6}$
for the ${3/2}^+$ state of $^{11}Be$.} \label{fig6}
\end{figure}

\section {Calculation of resonances}

\noindent For a particular resonance state having spin-parity
$J^{\pi}$, if there exists a bound state of the same $J^{\pi}$, we
start with the lowest lying state (ground state) wave function
$\psi_{0}(r)$ having energy $E_{0}$ and construct the isospectral
potential $\hat{v}(r;\lambda)=\hat{v}_{1}(r;\lambda)+E_{0}$
according to Eq.~(\ref{isovhat}). On the other hand if there is no
bound state with the same $J^{\pi}$, we solve Eq.~(\ref{isoeq}) for
a positive energy $E$ subject to the boundary condition
$\psi_{E}(0)=0$ and normalized to a constant (fixed) amplitude of
oscillation in the asymptotic region, to get $\psi_{E}(r)$ and
construct $\hat{v}(r;\lambda)$ according to Eqs. (\ref{ie}) and
(\ref{vhat}). In both cases $\hat{v}$ approaches $v$ for
$\lambda\rightarrow+\infty$ and develops a deep and narrow well
followed by a high barrier near the origin for
$\lambda\rightarrow0+$. The deep well and high barrier combination
effectively traps the system giving rise to a
quasi-bound state.\\

\noindent We calculate the probability of the system to be trapped
within this enlarged well-barrier combination,
\begin{equation}
 C(E)={\displaystyle\int}_{0}^{r_{B}} {[\hat{\psi}_{E}(r^{\prime})]}^{2} dr^{\prime}  \;,
\label{ce}
\end{equation}
where $r_{B}$ is the position of the top of the barrier. The
quantity $C(E)$ as a function of $E$ shows a resonance behaviour -
$C(E)$ has a maximum at the resonance energy $E=E_{R}$. This
resonance curve becomes very sharp for small positive values of
$\lambda$, and approaches a delta function for
$\lambda\rightarrow0+$. Although the resonance energy is independent
of $\lambda$, one has to make a judicious choice of $\lambda$, since
for extremely small values of $\lambda$, numerical errors in the
wave function within the extremely narrow well becomes very large. A
plot of probability $C(E)$ as a function of energy $E$, for the
system to be within the trapping potential according to
Eq.~(\ref{ce}), for $\lambda = 4\times 10^{-5}$ is shown in Fig. 2.\\

\noindent The width $\Gamma$ of the resonance is obtained from its
mean life $\tau$, using the time energy uncertainty relation. The
mean life is the reciprocal of the decay constant, which is a
product of the number ($n_{c}$) of impacts on the barrier and the
transmission probability ($T$). We estimate $n_{c}$ semiclassically
as the reciprocal of time of flight within the well of $\hat{v}$
between the classical turning points $a$, $b$ and $T$ is obtained by
the WKB approximation for the transmission through the barrier of
$\hat{v}$ with resonance energy $E_{R}$:
\begin{equation}
T= exp[-2 {\displaystyle\int}_{b}^{c}
\sqrt{\frac{2\mu}{\hbar^2}(\hat{v}(\lambda;r)-E_{R}})  dr]
\end{equation}
where $b$ and $c$ are the classical turning points of the barrier. The final
expression for $\Gamma$ is
\begin{equation}
\Gamma=2\:\:\sqrt{\frac{\hbar^2}{2\mu}}\:\:\frac{ exp(-2
{\displaystyle\int}_{b}^{c}
\sqrt{\frac{2\mu}{\hbar^2}(\hat{v}(\lambda;r)-E_{R}})  dr)}
{{\displaystyle\int}_{a}^{b} \frac{dr}{\sqrt{(E_{R} -
\hat{v}(\lambda;r))}}}. \label{gamma}
\end{equation}

\noindent We have verified by direct calculation that $\Gamma$ is
independent of $\lambda$ within numerical errors.\\

\section {Results and Conclusions}

\noindent Considering $^{11}$Be as a two-body system ($^{10}$Be +
$n$), we investigated the ${5/2}^+$, ${3/2}^-$ and ${3/2}^+$
resonant states. The original two-body interaction potential was
constructed microscopically a single folding approach using DDM3Y
effective interaction. Successful study of resonant states of weakly
bound nuclei ($^6$He, $^6$Li, $^6$Be) was previously carried out
using phenomenological bound state
potentials~\cite{SD03,SD04,SD042}. The previous study~\cite{MA09} of
${5/2}^+$ resonance state of $^{11}$Be was carried out using
Wood-Saxon potential for $^{10}$Be(core) - n potential. Parameters
were adjusted to reproduce the two bound states ${1/2}^+$ and
${1/2}^-$ and one resonance state ${5/2}^+$. Limitation of such a
choice of potential could be circumvented by the use of realistic
microscopic potential, which increases the scope of study to include
multiple resonant states as evident from our study. In the present
work, DDM3Y potential is used for the first time in the analysis of quasi-bound states of $^{11}$Be.\\

\begin{table}
{\bf Table 1:}\\

Comparison of the experimental results~\cite{SH04,CA01} with the
calculated values of the present work using isospectral potential
(IP) for the resonance states of $^{11}$Be.\\

\begin{tabular}{|l|l|l|l|l|}
\hline \multicolumn{1}{|l}{J$^{\pi}$}&
\multicolumn{1}{|l}{$E_x$~\cite{SH04,CA01}}&
\multicolumn{1}{|l}{$\Gamma$~\cite{SH04,CA01}}&
\multicolumn{1}{|l}{$E_x(IP)$}& \multicolumn{1}{|l|}{$\Gamma(IP)$}\\
&(MeV)&(keV)&(MeV)&(keV)\\
\hline
5/2$^+$&1.77 $\pm$ 0.02& 100 $\pm$ 10&1.77&124\\
3/2$^-$&2.67 $\pm$ 0.02& 200 $\pm$ 10&2.67&188\\
3/2$^+$&3.89 $\pm$ 0.02& $<$ 50&3.40&110\\
\hline
\end{tabular}
\end{table}

\noindent The bound state corresponding to the ground state of
$^{11}$Be [$^{10}$Be$({0}^{+})\bigotimes(s_{1/2})$] for $l=0$ and
$J^{\pi}= {1/2}^+$ is reproduced by normalizing the DDM3Y potential
by a factor of 1.5. This potential was used to construct the
isospectral potential $\hat{v}$ of Eq.~(\ref{vhat}) for all the
resonant states. We next study the ${5/2}^+$ resonance state of
$^{11}$Be [$^{10}$Be$({0}^{+})\bigotimes(d_{5/2})$]. Since the
ground state has a different spin-parity and there is no bound
${5/2}^+$ state, we use the BIC technique. The effective two-body
potential $v(r)$ for the ${5/2}^+$ state presents a shallow well
followed by a low barrier. Now we solve Eq.~(\ref{isoeq}) for energy
$E > 0$, subject to the boundary condition at $r=0$, viz.,
$\psi_{E}(0)=0$. Then $\psi_{E}(r)$ is normalized to a constant
(fixed) amplitude of oscillations in the asymptotic region. This
$\psi_{E}(r)$ is used to construct $\hat{{v}}(r;\lambda)$ according
to Eq.~(\ref{ie}) and (\ref{vhat}), choosing an appropriately small
value of $\lambda$, such that there is a narrow and a deep well
followed by a high barrier which effectively traps the system.  The
choice of $\lambda$ was made to ensure a sharp response of $C(E)$
for an optimal numerical calculation. The excitation energy $E_{x}$
is sum of the resonance energy $E_{R}$ and the ground state energy
of 0.5 MeV. As evident from Fig. 2, the excitation energy of
${5/2}^+$ state is at $E_{x}$ = 1.77 MeV which is same as the
experimental value~\cite{SH04}. The width is obtained from
Eq.~(\ref{gamma}) as $\Gamma=124$ keV, which slightly deviates from
the experimental finding ($\Gamma_{exp}=100\pm 10$ keV). We also
found that $\Gamma$ is independent of $\lambda$. Next we investigate
the ${3/2}^-$ resonant state at excitation energy 2.67 MeV . We
extend the BIC technique to calculate the ${3/2}^-$ resonance state
[$^{10}$Be$({0}^{+})\bigotimes(p_{3/2})$]. Since there is no
bound-state of same spin-parity, the same procedure of BIC is
employed. For appropriate choice of $\lambda$, deep-welled
isospectral potentials are constructed. The plots of
$\hat{v}(r;\lambda)$  corresponding to $ E=E_{x}$ for different
values of $\lambda$ are shown in Fig. 3 along with the original
folding potential $v(r)$. It should be noted that with the decrease
of $\lambda$ there is a dramatic increase in the minimum of the well
as well as the barrier height, both shifting towards the origin.
Classical turning points a, b and c are marked for the well-barrier
combination of $\hat{v}(r;\lambda)$ for $\lambda$ = 0.00001 in Fig.
3. Wave functions at the excitation energy of $ E_{x}=2.67 $ MeV for
${3/2}^-$ resonant state is shown in Fig. 4. The inset of Fig. 4
shows the wave function plot for $\lambda=10^{-3}$ in an expanded
scale up to 100 fm. All resonant state wave functions of Fig. 4 show
similar features but plots are shown up to 10 fm to highlight their
relative magnitudes within the trapping potential. This figure
illustrates the increase of trapping ability of isospectral
potentials $ \hat{v}(r;\lambda)$ with the decrease of $\lambda$
values. A plot of $C(E)$ as a function of $E$ for various $\lambda$
shows how the trapping effect of $\hat{v}(r;\lambda)$ increases as
$\lambda$ decreases (Fig. 5). As is evident from Fig. 5, the
sharpness of the peak increases rapidly as $\lambda\rightarrow 0+$.
It is apparent from the plots that there is a resonant state at
energy $E_{x}$ = 2.67 MeV. Width obtained from Eq.~(\ref{gamma}) is
$\Gamma=188$ keV. In a similar way, ${3/2}^+$ resonance state
[$^{10}$Be$({2}^{+})\bigotimes(s 1/2)$] is investigated using the
same procedure. Plots of probability $C(E)$ shown in Fig. 6 indicate
that there is a resonant state at $E_{x}$ = 3.40 MeV. Width $\Gamma
= 110$ keV was obtained semiclassically from Eq.~(\ref{gamma}).
Again we find the independence of the arbitrary parameter $\lambda$
for accurate determination of resonant state energy. Choice of
$\lambda$ only facilitates in detection of resonance states and
optimization of numerical computation. Excitation energy for the
${3/2}^+$ state differs slightly from the experimental value. The
experimental values~\cite{SH04} and our theoretical calculations are
given in Table 1. The study of resonance states of $^{11}$Be, by our
procedure clearly indicates the effectiveness of our technique in
handling resonance states of weakly bound nuclei.\\

\noindent D. Das acknowledges CSIR, India for financial support.\\

\end{document}